\documentclass[aps,reprint,amsmath,amssymb,superscriptaddress,showpacs,prb]{revtex4-2}
\usepackage{graphicx}

\begin{document}

\title{Qubits, entangled states, and quantum  gates realized on a set of classical pendulums}

\author{A.~V.~Nenashev}
\thanks{On leave of absence from Rzhanov Institute of Semiconductor Physics
  and the Novosibirsk State University, Russia}
\email{E-mail:  alexey.nenashev24@gmail.com}
\affiliation{Department of Physics and Material Sciences Center,
Philipps-Universit\"at Marburg, D-35032 Marburg, Germany}

\author{F.~Gebhard}
\affiliation{Department of Physics and Material Sciences Center,
Philipps-Universit\"at Marburg, D-35032 Marburg, Germany}

\author{K.~Meerholz}
\affiliation{Department f\"ur Chemie, Universit\"at zu K\"oln,
  Luxemburger Stra\ss e 116,
  50939 K\"{o}ln, Germany}

\author {S.~D.~Baranovskii}
\email{E-mail:  sergei.baranovski@physik.uni-marburg.de}
\affiliation{Department of Physics and Material Sciences Center,
  Philipps-Universit\"at Marburg, D-35032 Marburg, Germany}
\affiliation{Department f\"ur Chemie, Universit\"at zu K\"oln,
  Luxemburger Stra\ss e 116,
  50939 K\"{o}ln, Germany}


\begin{abstract}
Here we show that the concepts behind such terms as entanglement, qubits, quantum gates, quantum error corrections, unitary time evolution etc., which are usually ascribed to quantum systems, can be adequately realized on a set of coupled classical pendulums.
\end{abstract}

\maketitle   

\section*{Introduction}

Describing the basic concepts of quantum computing and quantum information, such as qubits, entanglement, quantum gates, quantum error corrections, unitary time evolution, Berry phase, etc., one usually addresses quantum systems, for instance, electrons, photons, atomic nuclei and so on  \cite{Stolze2008quantum,Nielsen2010quantum,Bouwmeester2013physics}. However, it is known that at least some of these basic concepts can be realized on classical systems. For instance, the Bloch sphere, which usually serves to illustrate qubit states, and the Berry phase is a representation of the classical Foucault pendulum \cite{Berry1990,Wharton2011classical,Linck2013}. The quantum-classical connection has been recently demonstrated for entangled photon anti-correlations, when analyzed via classical electromagnetic theory~\cite{Wharton2023}. The Bell inequality violations, often considered as the essence of quantum behaviour, were further demonstrated in some classical optics effects~\cite{Goldin2010,Qian2015,Song2015}.

The analogy between quantum and classical concepts has been mostly discussed with respect to the classical analogs of quantum entanglement
. One of such analogs was constructed using classical light beams mentioning that the analogy with quantum entanglement does not include nonlocality, the latter being considered of exclusive quantum nature~\cite{Spreeuw1998}.
Nonlocality is the key feature of the entanglement concept~\cite{Karimi2015}. If this feature lacks in classical realizations, the quantum-classical analogy is significantly limited.

In this work, we show that a classical system of coupled pendulums contains all above-mentioned features, including the feature of nonlocality. This calls for rethinking the limits of the quantum-classical analogy.

\section*{One qubit — two pendulums}

A qubit is a two-level system. In an attempt to visualize what a qubit is, we start from looking at a single energy level that has an energy $E$ and, therefore, a frequency $\omega = E/\hbar$. The complex amplitude $A$ of the probability that the system resides on this level is $A = | A |\exp {({\rm i}\varphi)}$, where $\varphi$ is a phase. This phase changes in time $t$ as $\varphi(t) = \varphi(0) + \omega t$, where $\varphi(0)$ is the initial phase. The occupation probability of this level is $| A | ^2$.

Let us represent this level as a classical oscillator with the frequency $\omega$. Henceforth, we use a pendulum to illustrate this oscillator. If we swing a pendulum to amplitude $| A |$, with the initial phase $\varphi(0)$, then its oscillations will symbolize the quantum state with the probability amplitude $A = | A |\exp {({\rm i}[\varphi(0) + \omega t])}$   of residing on the given energy level. The energy of the pendulum oscillations is proportional to $| A | ^2$, symbolizing the occupation probability of the given level.

Now we consider a qubit — a system with two basis states $\mid \downarrow \rangle$ and $\mid \uparrow \rangle$. If we want the qubit not to change its state in the course of time, we must set the energies of these states equal (i. e., have a twofold degenerate energy level). Hence, in our pendulum model, a qubit is represented as two equal pendulums, oscillating with the same frequency $\omega$. Notably, the amplitudes and the initial phases of their oscillations can be different. This defines the full variety of a qubit’s states.
Since the equations of motion of the pendulums are linear, the principle of superposition applies, i.e., if we sum up two different oscillations of the two-pendulum system, the resulting oscillation will also obey the equations of motion. In just the same way, one can sum up different states of a qubit, i. e., construct superpositions of states. Clearly, the analog of quantum superposition in our model is the summing up of such oscillations.

\subsection*{The Bloch sphere}
\begin{figure}
	\includegraphics[width=\linewidth]{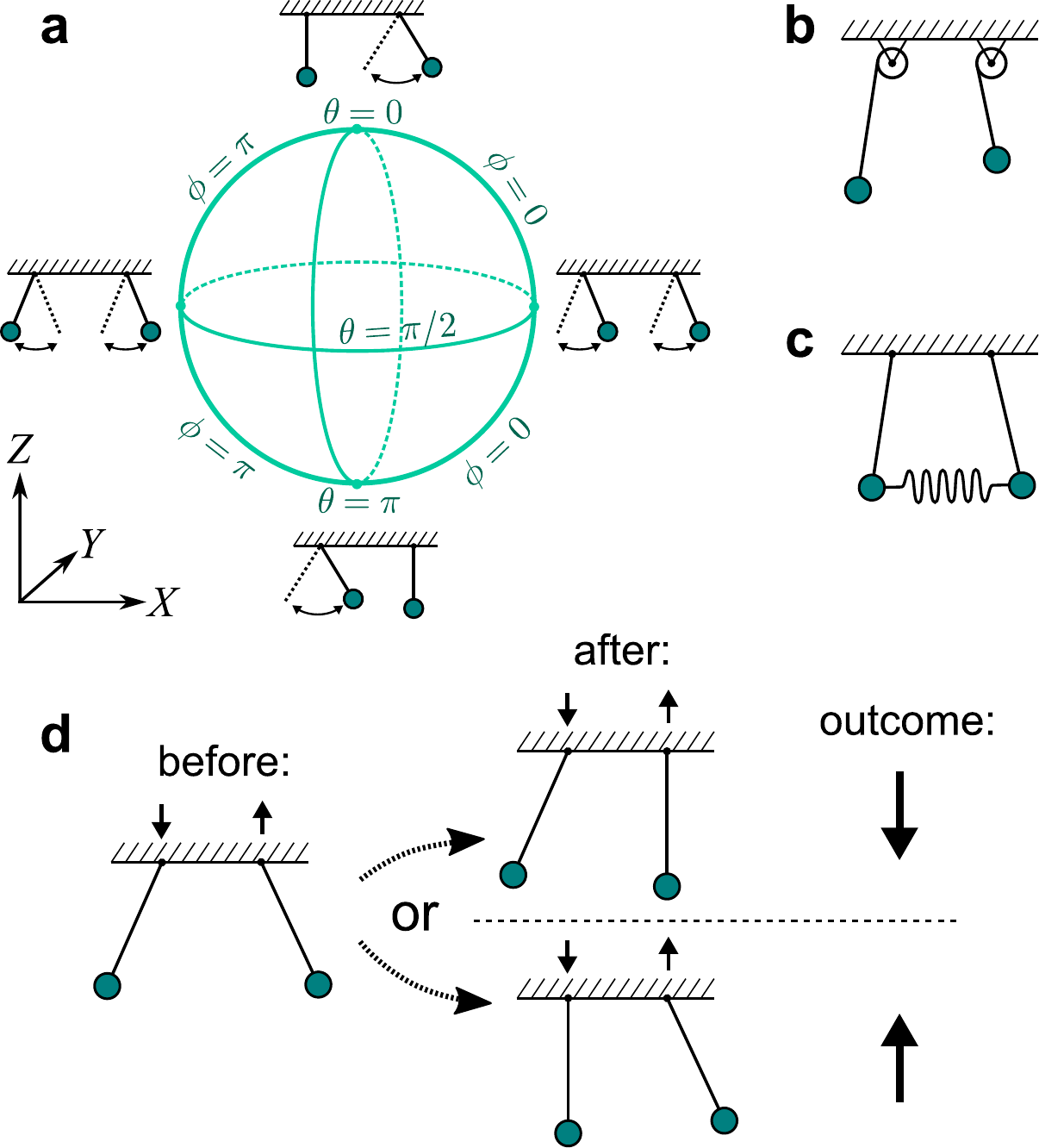}
	\caption{Representation of a qubit by a couple of pendulums. (a) Different oscillation modes on a Bloch sphere; (b) Control of frequencies for rotations around the $Z$-axis; (c) The pendulums coupled by a spring for rotations around the $X$-axis; (d) Measuring the qubit.}
	\label{fig:one_qubit}
\end{figure}

The sum of probabilities that the qubit is in state $\mid \downarrow \rangle$ and in state $\mid \uparrow \rangle$ is equal to unity. In our model, this means that the total energy of oscillations of two pendulums adds up to unity, $| A_{\downarrow} | ^2 + | A_{\uparrow} |^2 =1$, where $A_{\downarrow}$ and $A_{\uparrow}$ are the complex amplitudes of the two pendulums. Various modes of oscillations of the two-pendulum system differ from each other in two parameters: (1) how the energy is distributed among the pendulums, and (2) the difference between phases of the pendulum oscillations. It is convenient to represent these parameters as two coordinates on a sphere (a globe) — the latitude and the longitude. This is how we arrive at the Bloch sphere illustrated in Fig.~\ref{fig:one_qubit}a. The south pole of this sphere, state $\mid \uparrow \rangle$, represents the situation where the whole energy is given to the 1st pendulum, and the 2nd pendulum is at rest. The north pole, state $\mid \uparrow \rangle$, represents the situation where, vice versa, the 1st pendulum is at rest and the 2nd one oscillates.
The equator of the Bloch sphere represents those states, where the energy of oscillations is evenly distributed among the pendulums. The polar angle $\theta$ counts from the north pole and defines a parallel on the Bloch sphere. It relates to the amplitudes $| A_{\downarrow} |$ and $| A_{\uparrow} |$ as $| A_{\downarrow} | = \sin{(\theta/2)}$  and $| A_{\uparrow} | = \cos{(\theta/2)}$. The azimuthal angle (a longitude) $\varphi$, that defines a meridian, is just a difference between the oscillation phases of the pendulums.

When the frequencies of the two pendulums are equal to each other (i. e., the energy levels coincide), then the oscillation mode does not change in the course of time. For example, if the pendulums were moved in opposite phases, they would move so at any moment of time. It means that any state of the qubit is stationary. The flow of time does not change the state, up to multiplication by a phase factor $\exp({\rm i}\omega t)$. Concomitantly, a point on the Bloch sphere, which denotes this state, does not move in time.

\subsection*{One-qubit quantum gates with two pendulums}

Let us assume that we can change the frequencies of the pendulums during their oscillations. For example, let the pendulum’s thread be wound on a spool, and we can change the length of the pendulum’s leg (and hence the frequency) just by winding the thread on or off, as illustrated in Fig.~\ref{fig:one_qubit}b.

Let us realize with two pendulums a unitary transformation described by the two matrices $M_1 = \left(
\begin{array}{cc}
            1 & 0 \\
            0 & e^{{\rm i}\Delta \varphi} \\
\end{array}
\right)$ and $M_2 = \left(
\begin{array}{cc}
            e^{-{\rm i}\Delta \varphi/2} & 0 \\
            0 & e^{{\rm i}\Delta \varphi/2} \\
\end{array}
\right)$, which are equal to each other up to a common phase factor. To this end, let us increase the frequency of the 2nd pendulum by some value $\Delta\omega$, and, after the time interval $\Delta t$, return to the initial frequency. Then, the phase
difference between the pendulums will increase by the value $\Delta \varphi = \Delta\omega \cdot \Delta t$. Since the phase difference is nothing else
but the longitude on the Bloch sphere, the longitude will increase by $\Delta \varphi$. As a result, the whole Bloch sphere will rotate around the vertical
axis $Z$ by the angle $\Delta \varphi$. For example, if $\Delta \varphi = \pi$, this manipulation turns the pendulums oscillating with equal phases into pendulums oscillating with opposite phases, and vice versa. From the point of view of qubit states, these are the unitary transformations described by the two matrices, $M_1$, $M_2$ given above.

Next, we realize the NOT quantum gate with two pendulums. For this purpose, we must be able to rotate the Bloch sphere around the $X$-axis. To achieve this goal, we connect the pendulum’s weights with a spring (see Fig.~\ref{fig:one_qubit}c) during some time interval $\Delta t$. A system of two connected pendulums possesses two normal modes. One mode corresponds to the pendulums swinging in phase, while the other mode corresponds to the pendulums swinging in opposite phases. Let us denote the difference between the frequencies of these normal modes as $\Delta \omega$. Connecting the pendulums for the time interval $\Delta t$ leads to the rotation of the Bloch sphere by the angle $\Delta \varphi = \Delta\omega \cdot \Delta t$.
Consider, for instance, the case $\Delta \varphi = \pi/2$. If the first pendulum was initially at rest (the state at the north pole of the Bloch sphere), then the 2nd pendulum will “wake up” the 1st one and will transfer to it half of its energy during the time $\Delta t$. Therefore, the system’s state will move from the north pole to the equator of the Bloch sphere. This evolution is described by transformation matrix  $M_3 =\left(
\begin{array}{cc}
            1 & {\rm i} \\
            {\rm i} & 1 \\
\end{array}
\right)/\sqrt{2} \,$,
 corresponding to the rotation of the Bloch sphere around the $X$-axis. In the case $\Delta \varphi = \pi$, the pendulums’ amplitudes will change places during the manipulation. For example, the state at the north pole (the 1st pendulum is at rest while the 2nd one moves) turns into the state at the south pole (the 1st pendulum moves and the 2nd one is at rest), and vice versa. This is the NOT gate, which is described by the unitary matrix $M_4 = \left(
\begin{array}{cc}
            0 & 1 \\
            1 & 0 \\
\end{array}
\right)$.

The two examples show that any linear interaction of the pendulums provides two normal modes that correspond to the two opposite points on the Bloch sphere. The whole sphere rotates due to this interaction around the line that connects these points. The angular frequency $\Delta \omega$ of this rotation is equal to the difference between the frequencies of the normal modes.

One can generalize these considerations as follows. Any “retuning” of the dynamics of the two-pendulum system, which continues within a finite time interval and does not violate the linearity of the equations of motion, gives rise to some turn of the Bloch sphere. Its action on qubit’s states is described by a $2 \times 2$ unitary operator $M$. This is the one-qubit quantum gate in its general form.

\subsection*{Measuring a qubit}

The motion of pendulums, which represents the qubit’s quantum state, is somewhat similar to the Kantian thing-in-itself. The only information accessible to the observer (i. e., given to the macroscopic world) is the measurement outcome. Let us consider such a measurement in the framework of the pendulum model.

We are interested in the measurement of a qubit state in the basis $\mid \downarrow \rangle$, $\mid \uparrow \rangle$. Quantum measurement is a random (probabilistic) choice. In our setting, it is a choice of one of the two pendulums. The probability of choosing a given pendulum is proportional to the energy of its oscillations. If only one of the two pendulums is moving, then the choice is pre-determined: the moving pendulum will be chosen. This corresponds to the postulate of quantum theory: if the state of a quantum system is an eigenstate of the measured quantity, then the outcome of the measurement is determined. In other states, an outcome (a chosen value) is random. For example, if the two pendulums have the same amplitudes of oscillations, then both results of the choice are equally likely.

What happens at the moment of choice? First, the chosen outcome becomes known to the observer. This is the only available information about the quantum system. Second, only the chosen pendulum continues to move, the other pendulum stops (see Fig.~\ref{fig:one_qubit}d), and, herewith, a new state is prepared.

Since the energy of the pendulum corresponds to the probability, then the total energy of the two pendulums before the measurement is equal to unity. After the measurement, one of the pendulums is stopped, hence the total energy is decreased. Therefore, we have to decrease the unit of energy correspondingly, so that the energy of the system remains to be equal to unity.

\subsection*{Berry phase in the pendulum representation}

A qubit can be represented not only as a system of two pendulums, but also as a single pendulum with two degrees of freedom. Therefore, the Foucault pendulum, which has two independent modes of movement along two horizontal axes, is a natural representation of a qubit. Rotation of the Foucault pendulum's plane of oscillation due to the rotation of Earth is known to be a direct classical analogy of the quantum-mechanical Berry phase~\cite{Berry1990,Wharton2011classical,Linck2013}. For the sake of completeness, we briefly review this analogy in the Supplementary Materials, see Section S1.

\section*{Two qubits — four pendulums}

\begin{figure}
	\includegraphics[width=\linewidth]{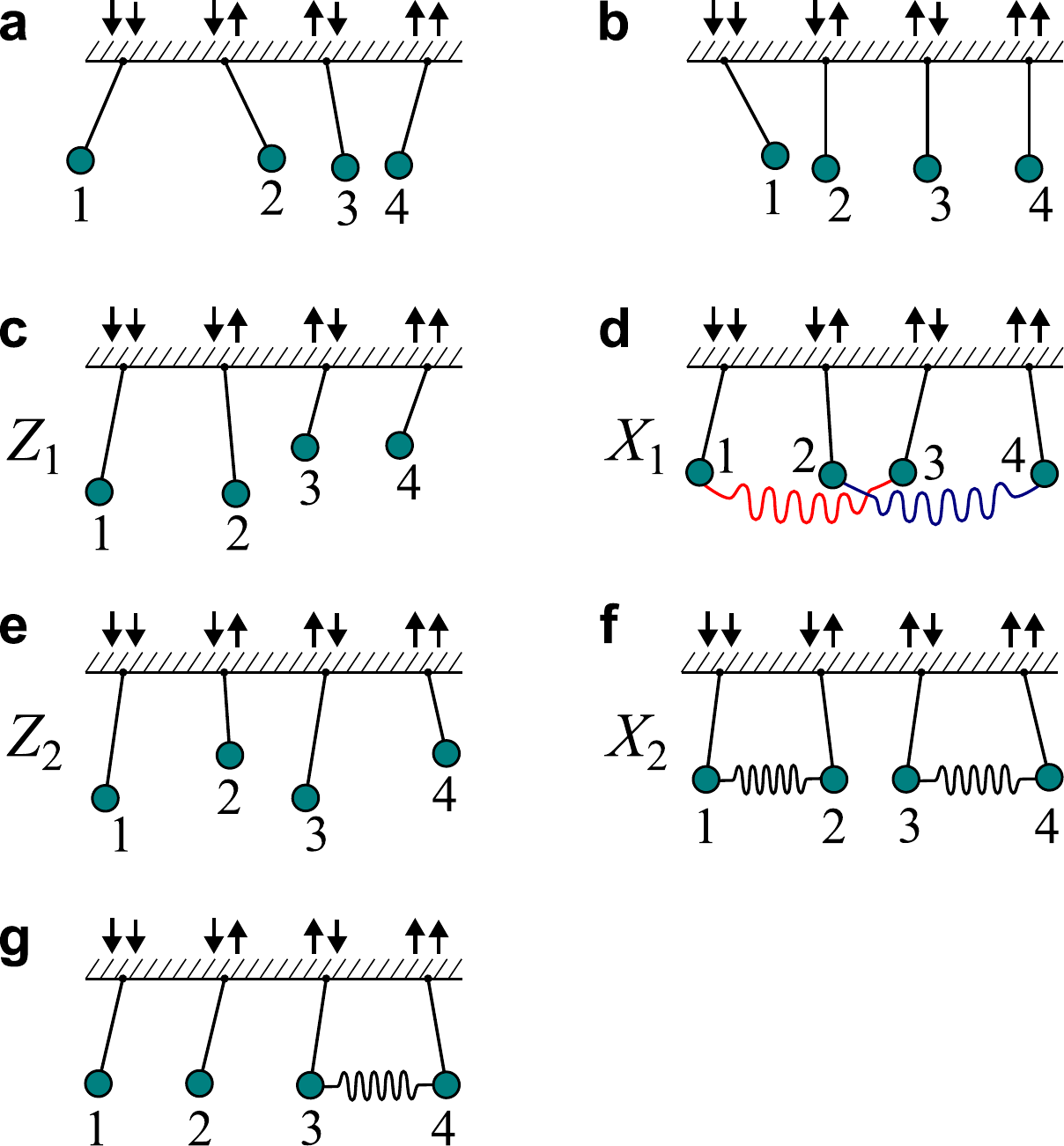}
	\caption{Representation of two qubits by four pendulums. (a) Example of possible oscillations; (b) The state after initialization; (c), (e) Control of frequencies for rotations around the $Z$-axis; (d), (f) Pendulums coupled by springs for rotations around the $X$-axis; (g) Pendulums coupled by a spring for rotation of the 2nd qubit controlled by the 1st qubit.}
	\label{fig:two_qubits_1}
\end{figure}

Since one qubit has 2 independent basis states $\mid \downarrow \rangle$ and $\mid \uparrow \rangle$, a system of two qubits has 2*2=4 basis states: $\mid \downarrow \downarrow \rangle$,  $\mid \downarrow \uparrow \rangle$, $\mid \uparrow \downarrow \rangle$, and $\mid \uparrow \uparrow \rangle$. Therefore, a system of two qubits is represented in our model by four pendulums, one pendulum per basis state, as illustrated in Fig.~\ref{fig:two_qubits_1}a. The total oscillation energy of the four pendulums is set to unity.

Let us formulate, using a system of four classical pendulums, the basic concepts of quantum computing, namely, initializing a two-qubit system, constructing one- and two-qubit quantum gates, and measuring a two-qubit system, as well as the concept of entanglement.

\subsection*{Initialization}

Let $\mid \downarrow \rangle$ denote the logical zero. Then, initialization is setting both qubits to zero and, hence, setting the whole system to the state $\mid \downarrow \downarrow \rangle$. Let us appoint the first pendulum to represent the basis state $\mid \downarrow \downarrow \rangle$, as illustrated in Fig.~\ref{fig:two_qubits_1}b. In order to perform the initialization, one should, therefore, swing the 1st pendulum to unit amplitude, and stop the other pendulums.

\subsection*{Quantum gates}
\subsubsection*{One-qubit quantum gates with four pendulums}

In order to perform a one-qubit operation with the 1st qubit, one should make the operation independent from the states of the 2nd qubit. The state of the 2nd qubit, $\mid \downarrow \rangle$ or $\mid \uparrow \rangle$,  depends on whether the pendulum’s number in Fig.~\ref{fig:two_qubits_1} is odd or even. Hence, one should warrant that operations with the odd pendulums in Fig.~\ref{fig:two_qubits_1} are performed in the same way as those with the even pendulums. For instance, consider rotations of the 1st qubit around the axes $Z$ and $X$. We denote these operations as $Z_1$ and $X_1$. As explained above for one-qubit systems, the rotation of a qubit around the axis $Z$ by the angle $\Delta \varphi$ is achieved by imposing a frequency difference $\Delta\omega$ between the involved pendulums for the time interval $\Delta t$, so that $\Delta\omega \cdot \Delta t = \Delta \varphi$. The difference between the pendulum frequencies can be achieved, for instance, by manipulating the pendulum length. Therefore, length difference between even pendulums in Fig.~\ref{fig:two_qubits_1}c is set the same as the length difference between the odd pendulums, aiming to rotate only the 1st qubit. Rotations around the $X$ axis are achieved by binding the pendulums with springs. In order to perform the rotation $X_1$, one should connect the odd and even pendulums in Fig.~\ref{fig:two_qubits_1}d with identical springs.

Similarly, in order to perform a quantum gate on the 2nd qubit, one should make the gate independent from the state of the 1st qubit. The state of the 1st qubit depends on whether the couple $\mid \downarrow \downarrow \rangle$ and $\mid \downarrow \uparrow \rangle$ (pendulums 1 and 2 in Fig.~\ref{fig:two_qubits_1}e), or the couple $\mid \uparrow \downarrow \rangle$ and $\mid \uparrow \uparrow \rangle$ (pendulums 3 and 4 in Fig.~\ref{fig:two_qubits_1}e) is considered.
The independence from the 1st qubit can be achieved by doing equal operations with the first and the second pairs of pendulums. In Figs.~\ref{fig:two_qubits_1}e and \ref{fig:two_qubits_1}f, rotations of the 2nd qubit around axes $Z$ and $X$ ($Z_2$ and $X_2$, correspondingly) are illustrated.

\subsubsection*{Two-qubit quantum gates}

Let us impose an interaction between qubits connecting pendulum 3 in Fig.~\ref{fig:two_qubits_1}g (basis state $\mid \uparrow \downarrow \rangle$) with pendulum 4 (basis state $\mid \uparrow \uparrow \rangle$) by a spring. Then, what happens with the second qubit depends on the state of the first qubit. In Fig.~\ref{fig:two_qubits_1}g, the rotation of the 2nd qubit is controlled by the 1st qubit. If the interaction durates for such time $\Delta t$ that $\Delta t \cdot \Delta \omega = \pi$, one gets the well-known “controlled NOT” (CNOT) quantum gate. Here $\Delta \omega$ denotes the difference between the frequencies of the normal modes of the two connected pendulums.
Other two-qubit operations can be modeled in a similar manner. For instance, if the frequency of the 4th pendulum is changed for some time, one gets the controlled-phase-change (CPHASE) gate. Furthermore, if pendulums 2 and 3 are connected by a spring during the time $\Delta t=\pi/\Delta \omega $, one gets the SWAP gate that swaps two qubits.

\subsection*{Qubit measurements with four pendulums}

\begin{figure}
	\includegraphics[width=\linewidth]{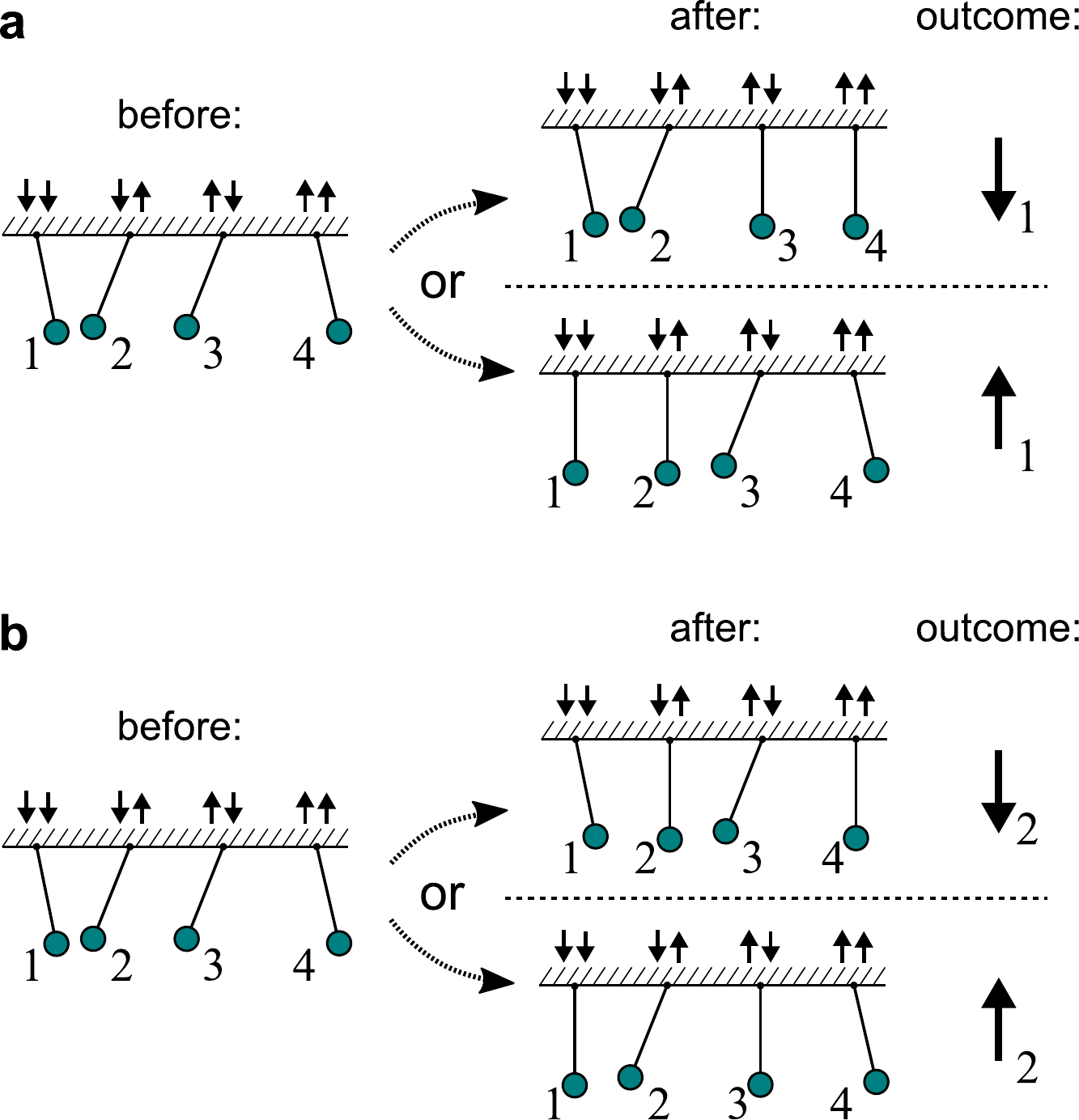}
	\caption{Independent measurements of the first qubit (a), or the second qubit (b), represented by the system of four pendulums.}
	\label{fig:two_qubits_2}
\end{figure}

In our model, the basis states $\mid \downarrow \rangle$ and $\mid \uparrow \rangle$ of the 1st qubit are related to the pair of pendulums 1 and 2 and to the pair of pendulums 3 and 4, respectively. The measurement of the 1st qubit in the basis $\{\mid \downarrow \rangle$, $\mid \uparrow \rangle\}$ is, therefore, the choice between the pairs of pendulums. The probability that a given pair will be chosen is equal to the fraction of the total energy that belongs to this pair. After the measurement, the chosen pair of pendulums continue to swing, and the other pair stops, as illustrated in Fig.~\ref{fig:two_qubits_2}a.
After the measurement, the 1st qubit occurs in a basis state $\mid \downarrow \rangle$, or $\mid \uparrow \rangle$, depending on the measurement outcome. This outcome (one bit of information) becomes available in the classical world. The total energy of those pendulums that continue to move becomes the new unit of energy.

The measurement of the second qubit in our pendulum model looks very similar to the measurement of the 1st qubit. The basis states $\mid \downarrow \rangle$ and $\mid \uparrow \rangle$ of the 2nd qubit distinguish between the odd (1 and 3) and the even (2 and 4) pendulums. Hence, the measurement of the 2nd qubit in the basis $\{\mid \downarrow \rangle$, $\mid \uparrow \rangle\}$ is the choice between the odd and the even pendulums, as illustrated in Fig.~\ref{fig:two_qubits_2}b.

When both qubits are measured, then the choice between the four possible outcomes $\mid \downarrow \downarrow \rangle$, $\mid \downarrow \uparrow \rangle$, $\mid \uparrow \downarrow \rangle$ and $\mid \uparrow \uparrow \rangle$ is performed. After this, only one pendulum (the chosen one) continues to move, and two bits of information become accessible to the classical world.

\subsection*{Entangled states}

A state of one qubit is defined by two parameters, for instance, by the latitude and the longitude on the Bloch sphere. In other words, the state space of a qubit is two-dimensional (it is on the surface of the Bloch sphere). Concomitantly, the state space of the two-qubit system is six-dimensional. These dimensions correspond to the real and imaginary parts of the complex amplitudes of four pendulums, excluding the common phase and the total energy, which are fixed.

The number of parameters of the whole two-qubit system (six) is larger than the net number of parameters (four) of its subsystems, qubits (2+2). Therefore, almost any state of the two-qubit system is not reduced to a union of states of separate qubits. This is the peculiarity of an entangled state.

Let us examine one of such entangled states, the singlet $| S \rangle = (\mid \downarrow \uparrow \rangle - \mid \uparrow \downarrow \rangle)/\sqrt{2}$  with the help of our pendulum model. In this state, the 2nd and the 3rd pendulums oscillate in opposite phases, while the other two pendulums are at rest.
Traditionally, qubits are discussed taking spins as example.  The basis states $\mid \downarrow \rangle$ and $\mid \uparrow \rangle$ are eigenstates of the $Z$-projection of the spin with eigenvalues $-1/2$ and $+1/2$. Any one-qubit quantum gate defines a rotation of the Bloch sphere. Quantum gates rotate the spin together with the Bloch sphere. This allows one to determine eigenstates of spin projections on any direction. Let us translate this observation to the language of pendulums.

When two qubits are in the singlet state, and the $Z$-projection of the 1st spin is measured (in the basis $\{\mid \downarrow \rangle$, $\mid \uparrow \rangle\}$, then only one of the pendulums will continue swinging after the measurement. It is either the 2nd pendulum or the 3rd one, depending on the measurement outcome. In both cases, the 2nd spin will occur in a state with a definite $Z$ projection, opposite to the measured $Z$-projection of the 1st spin. One can say that the measurement of the first qubit “\textit{acts at distance}” on the second one, bringing the latter into a definite state.

\begin{figure}
	\includegraphics[width=\linewidth]{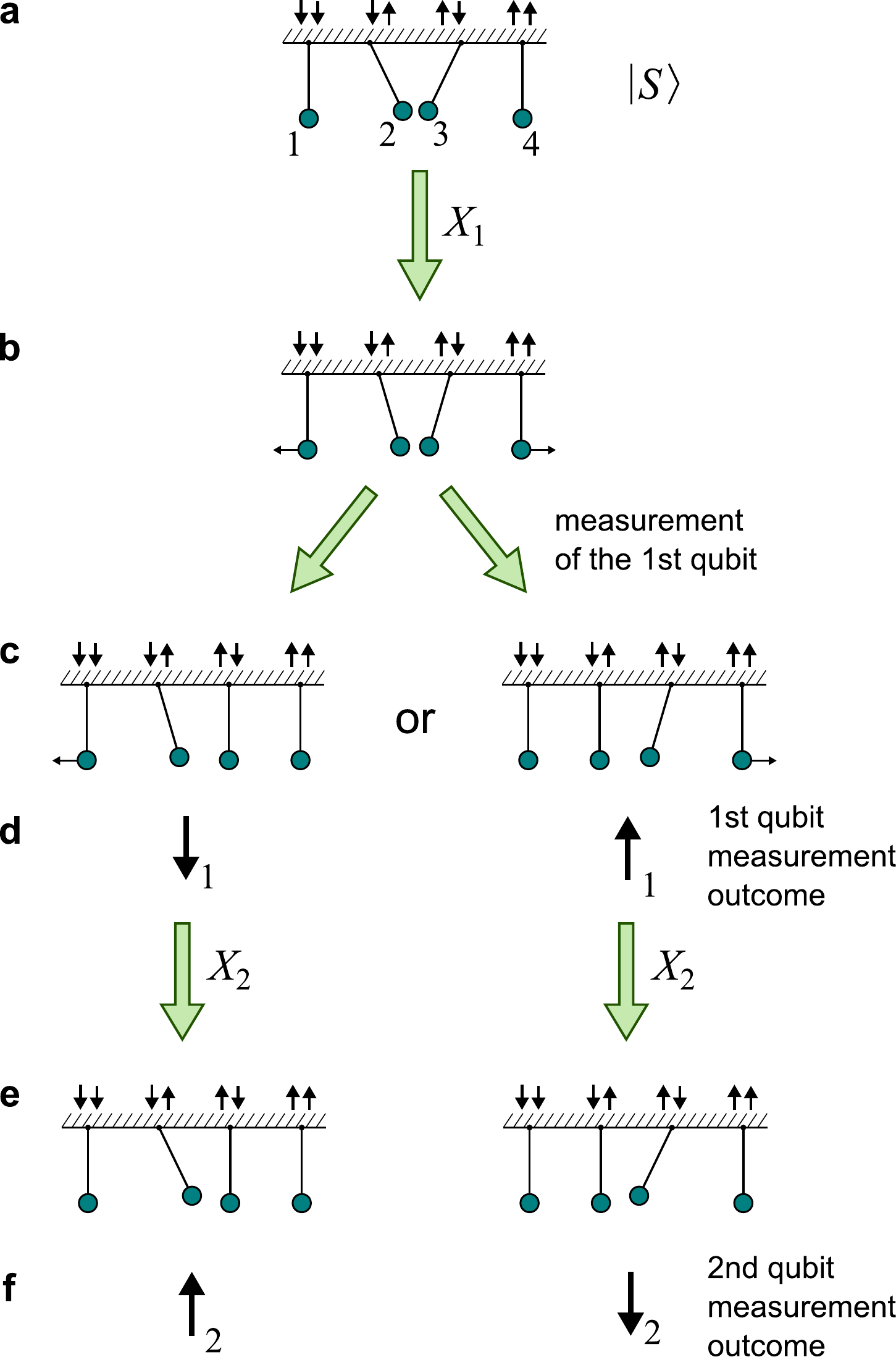}
	\caption{Consecutive measurements of qubits in entangled state of two qubits represented by four pendulums: (a) Singlet state $|S\rangle$; (b) The result of $X_1$ operation; (c) The state of the system after the measurement; (d) The two possible outcomes of the 1st qubit measurement; (e) Possible states after $X_2$ operations; (f) The two possible outcomes of the 2nd qubit measurement.}
	\label{fig:two_qubits_3}
\end{figure}

Let us discuss in more detail this \textit{action at distance} using the pendulum language for spins. For instance, let us consider what happens with the second spin after measuring the $Y$-projection of the first spin instead of its $Z$-projection. Measuring the spin projection on the $Y$-axis is equivalent to the spin rotation around the $X$-axis by $90^{\circ}$, which turns the $Y$-axis into the $Z$-axis, and the subsequent measuring the $Z$-projection. Let us perform such a measurement with the first qubit, starting from the singlet state $| S \rangle$ of the two-qubit system, as depicted in Fig.~\ref{fig:two_qubits_3}(a-d).

Just before measuring the $Z$-projection, we get the state of motion, in which all the pendulums have equal amplitudes, though different phases of their oscillations. Within the pair of pendulums 1 and 2, the phase difference is equal to $90^{\circ}$. The same phase difference, but with the opposite sign, appears in the pair 3 and 4. Therefore, before the measurement, the left and the right pairs of pendulums are exactly in the eigenstates of the $Y$-projection of the second spin, with opposite eigenvalues. The measurement chooses either the left or the right pair, and thus fixes the $Y$-projection of the second spin. The $Y$-projection of the second spin always turns out to be opposite to the measured $Y$-projection of the first spin. In order to verify this situation, we can continue our thought experiment and measure the $Y$-projection of the 2nd spin in the same way as it was done with the 1st spin above, as illustrated in Fig.~\ref{fig:two_qubits_3}. Clearly, the resulting value of the 2nd spin is always opposite to that of the 1st spin. This result generalizes to any other axis direction. Independently from the axis direction, the 2nd spin projection always occurs opposite to that of the 1st spin projection.

Entanglement in quantum systems is a phenomenon that “cannot be translated” into the classical language. Mathematically, this impossibility of ``translation'' is expressed by the fact that when measuring entangled states, Bell's inequality is violated, while in any classical analogue (like a pair of gloves), Bell's inequality cannot be violated.

Remarkably, the pendulum model in this article has a feature that allows for the violation of Bell's inequality and, consequently, for mimicking quantum entanglement. In fact, Bell's inequality holds for any theory that possesses local realism. However, our pendulum model is not local-realistic. Indeed, each pendulum corresponds to some basis state of the pair of qubits. If we consider the qubits as spatially separated, then each pendulum belongs to both qubits' locations simultaneously. That is, the pendulums are non-local in space. As a result, the pendulum model is realistic, but not local-realistic. Thus the violation of Bell's inequality is not forbidden. The full account of quantum entanglement is possible in our pendulum model due to non-local nature of the pendulums.

\section*{N qubits — $2^N$ pendulums}

\begin{figure}
	\includegraphics[width=\linewidth]{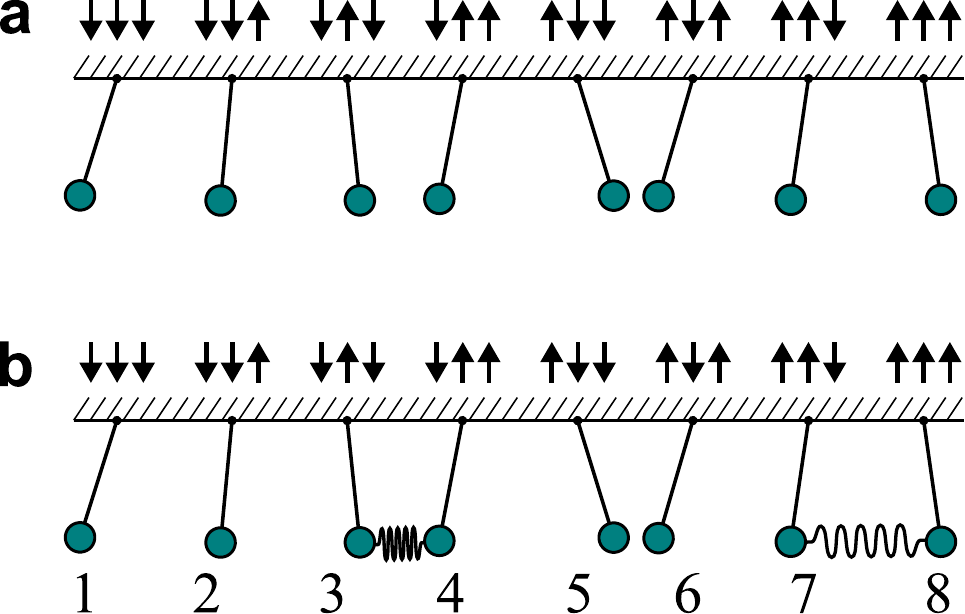}
	\caption{Representation of three qubits by eight pendulums. (a) Example of possible oscillations; (b) CNOT operation.}
	\label{fig:three_qubits}
\end{figure}

Following the same logic as for the one- and two-qubit systems, we can represent a system of $N$ qubits as a collection of $2^N$ pendulums. The picture below shows $2^3 = 8$ pendulums that describe a system of three qubits, as depicted in Fig.~\ref{fig:three_qubits}a.

The general principle is as follows. We take one pendulum per each basis state $\mid \downarrow \downarrow \downarrow \rangle$, $\mid \downarrow \downarrow \uparrow \rangle$, \ldots, $\mid \uparrow \uparrow \uparrow \rangle$. The motion of each pendulum is described by a complex amplitude, which is equal to a coefficient in the expansion of the quantum system’s wavefunction over the basis states.
All the manipulations with these pendulums are completely analogous to those described above for one or two qubits. For instance, the example of implementing the CNOT gate to 2nd and 3rd qubits in a three-qubit system is depicted in Fig.~\ref{fig:three_qubits}b.

The pendulum representation also naturally demonstrates the need for quantum error corrections during a quantum computing procedure. Briefly speaking, classical bits are protected from errors by their nonlinear dynamics. In contrast, the pendulums that represent the quantum bit do not possess any nonlinearity, and therefore inevitably must suffer from accumulating effect of perturbations. We consider this issue in the Supplementary Materials, see Section S2, where we provide a simple example of an error correction represented in the pendulum language.

\section*{Parallels between pendulums and qubits}

In conclusion, let us summarise the similarities between the collections of pendulums and the collections of qubits, listing the qubit features with their pendulum counterparts.
\begin{itemize}
  \item A separate qubit is a pair of pendulums (see Fig.~\ref{fig:one_qubit}).
  \item $N$ qubits are $2^N$ pendulums, in accord with the number of the degrees of freedom of a quantum system (see Figs.~\ref{fig:two_qubits_1}, \ref{fig:two_qubits_2}, \ref{fig:two_qubits_3} for $N=2$ and Fig.~\ref{fig:three_qubits} for $N=3$).
  \item The probability of a qubit state is the oscillation energy of the corresponding pendulum.
  \item Initialization of a qubit system in the state $|000...0\rangle$  is moving the 1st pendulum and keeping other pendulums at rest (see Fig.~\ref{fig:two_qubits_1}b for $N=2$).
  \item Quantum gates on a single qubit are linear modifications of pendulum's dynamics, such as varying the pendulum frequencies (see Fig.~\ref{fig:one_qubit}b) or coupling pendulums with springs (see Fig.~\ref{fig:one_qubit}c).
  \item One-qubit quantum gates on a system of two qubits are linear modifications of pendulum's dynamics, which involve either the pairs of even and odd pendulums (see Fig.~\ref{fig:two_qubits_1}c,d $\rightarrow$ manipulating the 1st qubit), or the 1st and the 2nd pairs of pendulums (see Fig.~\ref{fig:two_qubits_1}e,f $\rightarrow$  manipulating the 2nd qubit).
  \item A more general modification of linear dynamics, for instance, the one shown in Fig.~\ref{fig:two_qubits_1}g, provides a two-qubit quantum gate.
  \item Measuring of a qubit is controlled by probabilities of different outcomes. In the pendulum model, the probabilities are oscillation energies. After the measurement, only half of pendulums continue to oscillate, the rest pendulums are stopped. The selection between these two groups is controlled by the outcome value, as illustrated in Fig.~\ref{fig:one_qubit}d for the one-qubit setting and in Fig.~\ref{fig:two_qubits_2} for the two-qubit system.
\end{itemize}

These considerations clearly demonstrate that quantum systems can be, in many cases, described by classical pendulums, unlike believed so far. Whether our concept may lead to future applications, seems unlikely, considering the practical difficulty to couple a large number of pendulums, but remains to be seen.

\bibliography{scibib}

\section*{Acknowledgments}
A.N. thanks the Faculty of Physics of the Philipps Universit\"at Marburg
for the kind hospitality during his research stay. S.D.B. and K.M. acknowledge financial
support by the Deutsche
Forschungsgemeinschaft (Research Training Group ``TIDE'', RTG2591)
as well as by the key profile area ``Quantum Matter and Materials
(QM2)'' at the University of Cologne.

\section*{Supplementary materials}
Supplementary sections S1 and S2\\

\end{document}